\documentclass[aip,jcp,reprint]{revtex4-1}

\usepackage{graphicx}
\usepackage{gensymb}
\usepackage{color}
\usepackage{soul}

\begin{document}

\title{Transformation pathways in high-pressure solid nitrogen: from molecular N$_2$ to polymeric cg-N}

\author{Du\v{s}an Pla\v{s}ienka}
\email{plasienka@fmph.uniba.sk}
\affiliation{Department of Experimental Physics, Comenius University,
\\ Mlynsk\'{a} Dolina F2, 842 48 Bratislava, Slovakia}

\author{Roman Marto\v{n}\'{a}k}

\affiliation{Department of Experimental Physics, Comenius University,
\\ Mlynsk\'{a} Dolina F2, 842 48 Bratislava, Slovakia}


\date{\today}

\begin{abstract}

The transformation pathway in high-pressure solid nitrogen from N$_2$ molecular state to polymeric cg-N phase was investigated by means of \textit{ab initio} molecular dynamics and metadynamics simulations. In our study, we observed a transformation mechanism starting from molecular $Immm$ phase that initiated with formation of $trans$-$cis$ chains. These chains further connected within layers and formed a chain-planar state, which we describe as a mixture of two crystalline structures - $trans$-$cis$ chain phase and $planar$ phase, both with $Pnma$ symmetry. This mixed state appeared in molecular dynamics performed at 120 GPa and 1500 K and in the metadynamics run at 110 GPa and 1500 K, where the chains continued to reorganize further and eventually formed cg-N. During separate simulations, we also found two new phases - molecular $P2_1/c$ and two-three-coordinated chain-like $Cm$. The transformation mechanism heading towards cg-N can be characterized as a progressive polymerization process passing through several intermediate states of variously connected $trans$-$cis$ chains. In the final stage of the transformation chains in the layered form rearrange collectively and develop new intraplanar as well as interplanar bonds leading to the geometry of cg-N. Chains with alternating $trans$ and $cis$ conformation were found to be the key entity - structural pattern governing the dynamics of the simulated molecular-polymeric transformation in compressed nitrogen.

\end{abstract}

\maketitle

\section{Introduction}

High-pressure phases of nitrogen have been extensively studied due to dramatic changes occurring upon polymerization from the molecular state. For the uniquely strong bonding in N$_2$, polymeric single-bonded forms of N are prospective high energy density materials, in which a vast amount of energy may be potentially stored, either in form of pure nitrogen \cite{Uddin, Zarko} or as composite systems like CO/N$_2$ \cite{Raza} or alkali-metal azides \cite{Vaitheeswaran-Babu}. Though numerous polymeric phases of nitrogen were predicted theoretically, experimentally only two nonmolecular structures have been confirmed - the cubic gauche phase, first predicted by Mailhiot \textit{et al.} \cite{Mailhiot} and the layered polymeric phase possibly corresponding to the $Pba2$ structure predicted by Ma \textit{et al.} \cite{Ma}.

Nitrogen at low pressures and temperatures is a typical molecular solid and well-known N$_2$ phases include cubic $\alpha$-N$_2$, tetragonal $\gamma$-N$_2$ and rhombohedral $\epsilon$-N$_2$ \cite{Katzke-Toledano, Gregoryanz-N2-2002}. At higher temperatures, orientationally disordered forms appear, like hexagonal $\beta$-N$_2$, cubic $\delta$-N$_2$ and partially-disordered tetragonal $\delta\*$-N$_2$ \cite{Katzke-Toledano, Bini, Tassini-Gorelli-Ulivi, Stinton}. Other five observed molecular phases have yet undetermined symmetry \cite{Katzke-Toledano, Gregoryanz-zeta-N2, Hooper} - $\zeta$-N$_2$ \cite{Gregoryanz-N2-2002, Eremets-N-2004-2, Gregoryanz-N-2007}, $\zeta'$-N$_2$ \cite{Gregoryanz-N-2001}, $\kappa$-N$_2$ \cite{Gregoryanz-N-2007}, $\theta$-N$_2$ \cite{Gregoryanz-N2-2002} and $\iota$-N$_2$ \cite{Gregoryanz-N2-2002}. An extensive group-subgroup investigation of possible transformations between molecular phases was performed by Katzke and Tol\'{e}dano \cite{Katzke-Toledano}, and temperature-induced transitions in the molecular regime were studied in Refs. \cite{Katzke-Toledano, Presber, Temleitner}. Several molecular structures were predicted by computer simulations - e.g. N$_2$-N$_6$ \cite{Mattson-Sanchez-Portal}, $Immm$-N$_2$ \cite{Hooper, Caracas-Hemley}, $P42_12_12$-N$_2$ \cite{Pickard-Needs}, a structure composed of two different isomers of N$_8$ molecules \cite{Hirshberg} and many more \cite{Pickard-Needs, Hooper, Caracas-Hemley}.

Regarding polymeric nitrogen, the first nonmolecular form synthesized in the laboratory by Goncharov \textit{et al.} \cite{Goncharov-N-2000} is the amorphous $\mu$-phase with the proposed average coordination of 2.5. However, though it was pointed out that it is uncertain, whether this phase represents a genuine amorphous form or a fine mixture of different three-coordinated ($3c$) phases \cite{Goncharov-N-2000}. This narrow-gap semiconductor \cite{Gregoryanz-N-2001} was successfully brought down to ambient pressure when cooled to 100 K \cite{Eremets-N-2001}. Besides $\mu$-N, a possibly different (reddish) amorphous form was observed by Lipp \textit{et al.} \cite{Lipp}.

The first polymeric crystalline structure of N obtained experimentally is the cubic gauche phase - cg-N, which was first reported by Eremets \textit{et al.} in 2004 \cite{Eremets-N-2004-1, Eremets-N-2004-2} at pressure and temperature over 110 GPa and 2000 K. Later experiments managed to synthesize cg-N also at different conditions \cite{Gregoryanz-N-2007, Trojan, Lipp, Popov, Tomasino-2}. This insulating phase yields unusual structure with $I2_13$ symmetry, which is composed of fused N$_{10}$ rings connected in a way that maximizes the number of energetically favorable $gauche$ dihedral angles (of $lp$-N-N-$lp$ with $lp$ being the electron lone-pair), which minimize the effect of mutual $lp$ repulsion \cite{Mailhiot, Zahariev-N-2005}. In this conformation all N atoms are $3c$ and single-bonded and local N-N-N angles are nearly tetrahedral - cg-N may be therefore viewed within the $sp^3$-hybridization scheme with three out of five valence electrons occupying three single-bond orbitals and remaining two accommodating one lone-pair \cite{Chen-Fu-Podloucky}. Properties of cg-N were extensively investigated \cite{Yu-N-1, Yu-N-2, Yu-N-3, Zhao-N, Caracas} and it was found that this phase is a wide-gap optical material \cite{Yu-N-1} and should be metastable at ambient conditions \cite{Mailhiot, Barbee-N, Zhang-N}. In 2014, the second experimentally obtained polynitrogen form denoted as layered polymeric structure - LP-N was reported by Tomasino \textit{et al.} \cite{Tomasino-2}. This phase was synthesized along cg-N and amorphous N above 125 GPa and probably corresponds to the predicted orthorhombic $Pba2$ structure \cite{Ma}, which is formed by fused N$_7$ rings arranged in layers. The thermodynamic transformation pressure between cg-N and molecular $\epsilon$-N$_2$ is predicted to be only around 60 GPa at 0 K, but at finite temperature it shifts to considerably higher pressures, because molecular phases gain entropy much faster than cg-N upon the increase of temperature \cite{Erba}. However, there still remains a large discrepancy between the experimental and theoretical pressure of transition between N$_2$ and cg-N at 2000 K, which can be related to the kinetics of the transformation requiring breakage of extremely strong triple bonds in N$_2$ \cite{Erba}.

A large number of high-pressure nitrogen phases has been predicted from first-principles by various simulation techniques including evolutionary algorithms, random structure searching methods or simple structural optimizations. Proposed phases are either chain-like, planar or fully extended with various types of covalent networks. Two-coordinated ($2c$) chain structures include $ch$-phase \cite{Mailhiot} with alternating $trans$ and $cis$ dihedral conformations (see Fig.~\ref{chains} (a) for $trans$-$cis$ chain picture) and structures made out of zig-zag (zz) chains aligned in $Imma$ \cite{Alemany-Martins} or $Cmcm$ \cite{Mattson-Sanchez-Portal} structures. More extended phases include fully single-bonded, yet layered forms of black phosphorus (BP) and $\alpha$-arsenic (A7) composed of fused N$_6$ rings in chair conformation \cite{Mailhiot}. Other similar phases with N$_6$ rings in boat conformation include layered boat (LB) structure with $P2_1/m$ symmetry \cite{Zahariev-N-2005} and $Pnma$ zigzag sheet (ZS) phase \cite{Hu} \footnote{In Ref. \cite{Zahariev-N-2005}, it was recognized that the internal energy of single-bonded polynitrogen phases is related to the proportion of various $lp$-N-N-$lp$ dihedral angles. The minimal energetic configuration yields the $gauche$ conformation followed by second $trans$ minimum and $cis$ maximum \cite{Zahariev-N-2005}. The cg-N phase with all-$gauche$ angles thus yields lowest energy.}. By a systematic searching method 26 new metastable phases of nitrogen were discovered \cite{Zahariev-N-2006} demonstrating that a vast number of various structures are in fact feasible for element 7. By recognizing the helical structure motif in several polynitrogen forms \cite{Zahariev-N-2007} \footnote{Helical chains are stable in sulfur and selenium, which have one more valence electron, but in nitrogen they acquire stability only if connected \cite{Zahariev-N-2007}.}, rhombohedral chaired web (CW) phase was found \cite{Zahariev-N-2007} containing N$_6$ rings in chair conformation, which form voids that accommodate electron lone-pairs and thus reduce the repulsive energy. Other predicted polymeric phases of N include layered $Cmcm$ phase \cite{Caracas-Hemley}, fully three-dimensional $C2/c$ phase \cite{Oganov-Glass-USPEX}, layered mixed two-three-coordinated ($2c$-$3c$) $Pmna$ form \cite{Wang-N-2007} made up of fused N$_{10}$ rings, $P\bar{1}$ \cite{Yao} composed of puckered N$_8$ rings, layered $P\bar{4}2_1m$ \cite{Pickard-Needs} and $Cccm$ \cite{Wang-N-2013}, possibly superconducting $Pnnm$ \cite{Wang-N-2013} or helical tunnel $P2_12_12_1$ phase \cite{Ma, Pickard-Needs} with fused N$_4$ and N$_8$ rings. Several comparative studies were performed in order to discriminate relevancy of individual predicted polymorphs \cite{Kotakoski-Albe, Wang-N-2010} and it was concluded that most of them are metastable at all pressures and some even mechanically and/or dynamically unstable at certain conditions \cite{Kotakoski-Albe, Wang-N-2010}. Another very interesting and complex structure, surprisingly predicted to be stable at higher pressures is the diamondoid $I\bar{4}3m$ phase \cite{Wang-N-2012} with lattice sites occupied by N$_{10}$ tetracyclic cage-like molecules. At multi-TPa pressures, other unexpected phases with strong ionic character were predicted to be stable as well - all-nitrogen metallic salt $P4/nbm$, modulated form $P2_1$ and $R\bar{3}m$ and $I4_1/amd$ phases \cite{Sun-N}.

Nitrogen is isovalent to phosphorus, in which a first-order liquid-liquid transition (between P$_4$ molecular and P$_n$ polymeric liquids) was experimentally demonstrated for the first time among pure elements \cite{Katayama-P, Monaco-P}. This naturally led to an effort aimed at finding a similar phenomenon in nitrogen and first indications came from shock-compression experiments in fluid phase \cite{Radousky, Ross-Rogers} further supported by first-principles simulations along Hugoniots \cite{Mazevet}. While \textit{in situ} experiments led to conflicting results about the position of the melting line maximum and its interpretation \cite{Mukherjee-Boehler, Goncharov-N-2008, Tomasino-1}, recent \textit{ab initio} calculations supported the existence of structural \cite{Boates-Bonev-N-2009, Donadio} and electrical \cite{Boates-Bonev-N-2011} transition between molecular and chain-like liquids in N.

Current open questions concerning high-pressure behavior of nitrogen include structure of five molecular phases occurring above 60 GPa ($\zeta, \zeta', \kappa, \theta, \iota$), possible existence of some intermediate (meta)stable phases and knowledge of the transformation mechanism leading to cg-N. In this paper, we primarily focus on resolving the last problem. In the work of Zahariev \textit{et al.} \cite{Zahariev-N-2005}, it was suggested that transformations to layered structures might proceed via zig-zag chain geometry pattern, but it was also pointed out that $trans$-$cis$ chains might be involved instead, as far as many predicted polymeric forms of N can be viewed as different connections of either zig-zag, or, alternatively, $trans$-$cis$ chains - e.g. BP, A7, LB or ZS (see Fig.~3 of Ref. \cite{Zahariev-N-2005}). Although many new forms of nitrogen have been predicted in the last years, no \textit{ab initio} dynamical simulations have been performed to study the process of the molecular-nonmolecular transition - except for one study using classical force field \cite{Nordlund} and an investigation of an opposite process of cg-N shock-induced depolymerization studied by first-principles dynamic simulations \cite{Mattson-Balu, Beaudet-Mattson-Rice}. Here, we therefore present results of our \textit{ab initio} molecular dynamics and metadynamics study performed at pressures 110-120 GPa and temperatures 1000-2500 K and propose a transformation mechanism starting from an orthorhombic molecular phase that proceeds via a mixture of a $trans$-$cis$ chain phase and a $planar$ phase to the final cg-N. The properties of all observed intermediate phases and transformation mechanisms between them are discussed in detail.

\section{Simulation methods}

For electronic structure calculations, we used density functional theory based code VASP 5.3 \cite{VASP-1, VASP-2} employing projector augmented wave pseudopotentials and Perdew-Burke-Ernzerhof (PBE) parametrization of the exchange-correlation functional \cite{PBE}. Hard pseudopotential with outermost cutoff radius of 0.582 \AA \,was chosen in order to accurately describe molecular nitrogen with short N$_2$ dimers. The plane-wave energy cutoff was set to 900 eV.

Molecular dynamics (MD) simulations at constant pressure-temperature conditions ($NPT$ ensemble) were carried out using the Parrinello-Rahman barostat \cite{PR-2}, which is in VASP 5 implemented in combination with the stochastic Langevin thermostat. We also ran several metadynamics simulations \cite{metadynamics-1, cell-metadynamics-1, cell-metadynamics-3}, which are based on exploration of the energy landscape in the collective variables (or order parameters) space. In our simulations, we employed the version of metadynamics, which uses six independent parameters of the simulation supercell $h_{ij}$ as order parameters \cite{cell-metadynamics-1, cell-metadynamics-3} and is well-suited to study pressure-induced structural phase transitions. Concerning the choice of Gaussian parameters (width $\delta s$ and height $W$) we followed the relation $W=\delta s^2$, proposed in Ref. \cite{cell-metadynamics-3}. The specific values of $\delta s$ used in the simulations are provided in section III. All MD and most metadynamics simulations were performed on a sample of 144 atoms, for which the $\Gamma$-point sampling of the Brillouin zone provides sufficient accuracy for dynamical study of non-metallic phases.

\section{Results}

We now present results of our MD and metadynamics simulations and describe briefly the observed structural objects and phases, which are discussed in more detail later.

\subsection{Starting molecular phase $Immm$}

\begin{figure}[h]
\includegraphics[width=\columnwidth]{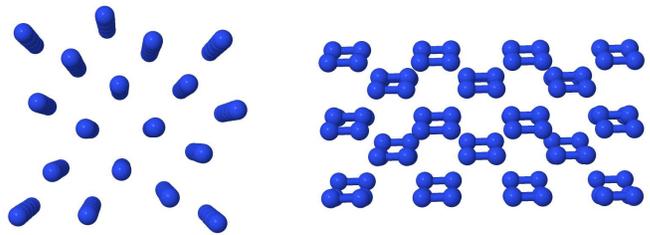}
\caption{$Immm$ molecular phase optimized at 110 GPa in (001) and (100) projection of the unit-cell orientation from Table~\ref{table} - left and right pictures, respectively. This phase was taken as the starting point for most of our simulations.}
\label{mol1}
\end{figure}

The \textit{NPT} MD simulations and most of our metadynamics runs were initiated from the $Immm$ molecular structure - Fig.~\ref{mol1} and Table~\ref{table}.
Natural choice for the starting structure in our study would be the high-pressure molecular phase $\zeta$ (or the lower pressure phase $\epsilon$). However, at high pressures, there are experimental difficulties in determining structure of low-$Z$ materials from XRD spectra and for that reason the symmetry of $\zeta$-N$_2$ still remains undetermined \cite{Gregoryanz-zeta-N2}. Despite an extensive theoretical effort \cite{Katzke-Toledano, Hooper}, $\zeta$-N$_2$ structure was up to now only proposed in Refs. \cite{Eremets-N-2004-2, Gregoryanz-N-2007, Katzke-Toledano, Hooper}. For the starting structure, we therefore chose to take the data provided for $\zeta$-N$_2$ in Ref. \cite{Eremets-N-2004-2} as the only available complete structural data suggested from the experiment. After structural optimization at several pressures below and above 100 GPa, the proposed $P222_1$ structure transformed into a different molecular phase, where all molecules were parallel to each other (Fig.~\ref{mol1}). For this phase, we found orthorhombic $Immm$ symmetry with unit-cell parameters and atomic positions (as well as density and PBE bandgap) at 110 GPa given in Table~\ref{table} (for \textit{molecular} A phase). We took this phase as the starting point for our MD and most metadynamics simulations, which were performed on a sample of 144 atoms (72 N$_2$ molecules) constructed as the $3\times3\times4$ supercell of the cubic $Immm$ unit cell.

This molecular phase was found earlier by Hooper \textit{et al.} \cite{Hooper} as one of the several candidates for the $\zeta$-N$_2$ structure - denoted as B1 in Ref.~\cite{Hooper}, for which the structural data corresponds to our $Immm$ phase. Also, molecular phase identified with the same spacegroup and with all molecules parallel to each other was described before by Caracas and Hemley \cite{Caracas-Hemley}. In their work, parameters for the $Immm$ phase at 90 GPa are - unit-cell vector lengths $a=$ 3.079 \AA, $b=$ 2.563 \AA, $c=$ 3.441 \AA, N$\equiv$N bond length = 1.092 \AA \,and N...N nearest intermolecular distance = 2.322 \AA. At the same pressure, values for our $Immm$ phase yield $a=$ 2.936 \AA, $b=$ 2.662 \AA, $c=$ 3.447 \AA, N$\equiv$N = 1.101 \AA \,and N...N = 2.077 \AA. Though there are some differences in exact values of certain parameters, the phases share the same structural character.

\begin{table*}
\centering
\begin{tabular}{|l|c|l|l|c|c|c|c|}

\hline

phase & symmetry & unit-cell parameters & $Z$ & Wyckoff and atomic positions & density & PBE bandgap \\
(status) &  & vector lengths; angle [\AA; $\degree$] &  &  & [g.cm$^{-3}$] & [eV] \\ \hline

\textit{molecular} A - & $Immm$ ($\#71$) & $a=$ 2.766 & 4 & \, \, 4j \, \, \, $\frac{1}{2}$ \, \, \, \, \,\, 0 \, \, \, \, \, \, 0.3364 & 3.705 & 0.37 \\
starting structure & orthorhombic & $b=$ 2.685 & &  &  &  \\
(see also Refs.~\cite{Hooper, Caracas-Hemley}) & & $c=$ 3.380 &  &  &  &  \\ \hline

\textit{molecular} B & $P2_1/c$ ($\#14$) & $a=$ 7.236; \, $\beta=$ 98.74 & 12 & \, \, 4e \, \, \, 0.3947 \, 0.3621 \, 0.8358 & 3.683 & 1.97 \\
(new) & monoclinic & $b=$ 2.588 & & \, \, 4e \, \, \, 0.9415 \, 0.6372 \, 0.9826 &  & \\
 & & $c=$ 4.092 & & \, \, 4e \, \, \, 0.2749 \, 0.6372 \, 0.8039  &  & \\ \hline

$trans$-$cis$ chain & $Pnma$ ($\#62$) & $a=$ 4.967 & 8 & \, \, 8d \, \, \, 0.9104 \, 0.9278 \, 0.8324 & 4.175 & \textit{semimetal} \\
(new) & orthorhombic & $b=$ 3.435 & & &  &  \\
 &  & $c=$ 2.612 &  &  &  & \\ \hline

$2c-3c$ chain-like & $Cm$ ($\#8$) & $a=$ 4.391; \, $\beta=$ 145.12 & 12 & \, \, 4b \, \, \, 0.4130 \, 0.8563 \, 0.2052 & 4.157 & 0.37 \\
(new) & monoclinic & $b=$ 7.759 & & \, \, 4b \, \, \, 0.4171 \, 0.8556 \, 0.5621 &  & \\
 &  & $c=$ 3.445 & & \, \, 2a \, \, \, 0.6624 \, 0 \, \, \, \, \,\, 0.3360 &  & \\
 &  &  &  & \, \, 2a \, \, \, 0.6777 \, 0 \, \, \, \, \,\, 0.9630 &  & \\ \hline

$planar$ & $Pnma$ ($\#62$)  & $a=$ 5.211 & 8  & \, \, 4c \, \, \, 0.1684 \, $\frac{3}{4}$ \, \, \, \, \,\, 0.1712 & 4.522 & 0.85 \\
(see also Ref.\cite{Hu}) & orthorhombic  &  $b=$ 2.203 & & \, \, 4c \, \, \, 0.1774 \, $\frac{3}{4}$ \, \, \, \, \,\, 0.5586 &  & \\
 &  & $c=$ 3.584 & &  &  & \\ \hline

\end{tabular}
\caption{Structural data, densities and PBE bandgaps of all proposed phases - molecular $Immm$ and $P2_1/c$, chain $trans$-$cis$, chain-like $2c$-$3c$ and $planar$-N (ZS), all at 110 GPa.}
\label{table}
\end{table*}

\subsection{Results - molecular dynamics}

The \textit{NPT} MD simulations were performed at 120 GPa and started at temperature of 1000 K, where $Immm$ molecular phase survived. The temperature was thereafter independently increased to 1300 K and to 1500 K and in both cases polymerization took place.

\begin{figure}[h]
\includegraphics[width=\columnwidth]{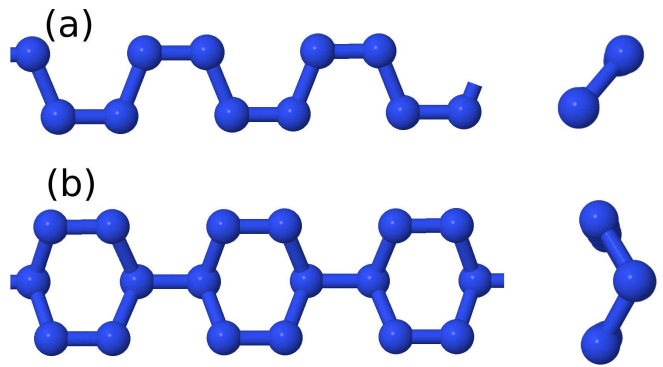}
\caption{Individual chains - fully $2c$ $trans$-$cis$ chain (a) and mixed $2c$-$3c$ chain-like polymer (b) from two views - right pictures correspond to projections along the chain axis.}
\label{chains}
\end{figure}

\begin{figure}
\includegraphics[width=\columnwidth]{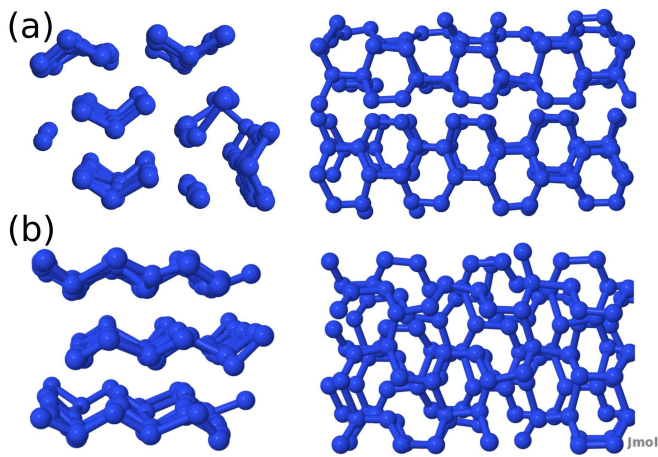}
\caption{(a) MD simulations at 120 GPa and 1300 K - the final state showing a mixture of mostly $2c$-$3c$ chains along one $trans$-$cis$ chain and some N$_2$ molecules. All chains and molecules are parallel to each other, while their orientation is random. (b) Final state from MD at 1500 K, which can be described as a mixture of two crystalline phases - the $trans$-$cis$ chain phase and the $planar$-N, that were both identified from this MD trajectory and other metadynamics runs. Nitrogen atoms have either two or three bonds and individual sheets are hence not completely polymerized. Left pictures correspond to (001) projections and right views are (010) projections of the simulation supercell - 3$\times$3$\times$4 unit-cell of $Immm$. Pictures were generated with Jmol \cite{Jmol}.}
\label{MD}
\end{figure}

At 1300 K, $trans$-$cis$ chains - Fig.~\ref{chains} (a) started to form, some of which further continued to merge into $2c$-$3c$ chain-like objects - Fig.~\ref{chains} (b). This process eventually resulted in a configuration consisting of $trans$-$cis$ and $2c$-$3c$ chains, which remained parallel to each other, though rotation around their axis was random (with no proper crystalline ordering). Some molecules also survived at this temperature - Fig.~\ref{MD} (a).

At 1500 K, polymerization also started with formation of $trans$-$cis$ and $2c$-$3c$ chains, which, however, continued to merge within layers until a quasi-planar state of partially connected chains was created - Fig.~\ref{MD} (b). From the structural character of this configuration, we identified two independent periodic arrangements based on geometry of $trans$-$cis$ chains - $trans$-$cis$ chain phase formed by individual disconnected chains and $planar$ phase made as interconnections of chains within sheets. Structural parameters of these two phases are given in Table~\ref{table}. This mixture of $trans$-$cis$ phase and $planar$-N then persisted on further compression and heating up to 200 GPa and 2500 K. As no other collective movements of atoms leading towards cg-N took place, MD trajectories ended up stuck in this layered state, although some interplanar bonds were occasionally formed throughout the run.

\subsection{Results - metadynamics}

\begin{figure}
\includegraphics[width=\columnwidth]{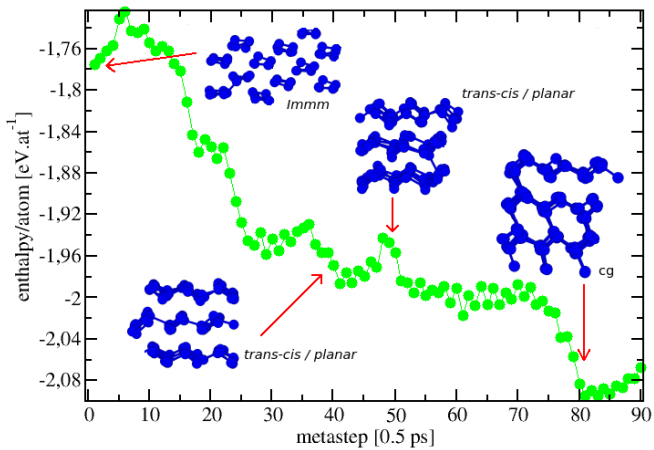}
\caption{Enthalpy per atom during metadynamics at 110 GPa and 1500 K with  $\delta s$ = 50 (kbar.\AA$^3$)$^\frac{1}{2}$ from metastep 1 to 91 (of time 0.5 ps). The starting $Immm$ molecular phase polymerizes making $trans$-$cis$ chains (around metastep 20), which further start to make some connections within planes and form a state that can be best described as a mixture of $trans$-$cis$ chain phase and $planar$ phase (between metasteps 25-50). Snapshots with mostly $trans$-$cis$ chain character at metastep 41 and with mostly $planar$-N character at metastep 49 are shown to illustrate states with different relative abundance of pure chain and planar phases. Note that enthalpy of metastep 49 configuration is slightly higher than enthalpy of metastep 41 state, which is consistent with relative enthalpies of pure $trans$-$cis$ chain phase and $planar$-N structures at 110 GPa (see Fig.~\ref{enthalpies}). This metastable mixed state then starts to rearrange and develop some interplanar bonds (after metastep 50), which finally leads to cg-N (starting to appear at metastep 80).}
\label{meta-enth}
\end{figure}

In our metadynamics simulations performed at 110 GPa and 1500 K with Gaussian width $\delta s$ = 50 (kbar.\AA$^3$)$^\frac{1}{2}$, on the other hand, we were able to follow complete transformation mechanism leading to cg-N within 80 metasteps. As shown e.g. in Ref. \cite{cell-metadynamics-3}, the metadynamics approach is able to reconstruct structural transformations passing through metastable intermediate states. The transformation started very similarly as in the MD at 1500 K - with creation of $trans$-$cis$ chains followed by formation of intermediate structure between $trans$-$cis$ chain phase and $planar$ phase (similarly as in Fig.~\ref{MD} (b)). However, in this case the transformation continued by bond reorganization inside the layers and by development of interplanar bonds in a process creating cg-N. The evolution of enthalpy in this metadynamics run is depicted on Fig.~\ref{meta-enth} along illustrations of the corresponding configurations - starting $Immm$ phase, mixed $trans$-$cis$ chain and $planar$-N states and the deepest enthalpy minimum corresponding to the appearance of high-temperature cg-N.

We also ran several other metadynamics simulations at different conditions and employing various values of $\delta s$ and also starting from the molecular $\epsilon$-N$_2$ phase and from some partially-polymeric structures. They all got stuck in various stages of polymerization, but never led to cg-N as the above described metadynamics run. In these other simulations, however, we found two different new phases - molecular $P2_1/c$ and $2c$-$3c$ chain-like, that are also described in Table~\ref{table}. The presence of a number of possible polymeric phases and the existence of several transformation pathways reflect the complex nature of the energy landscape of nitrogen that has to be traversed on the way from molecular to polymeric N.

Next, we describe all phases identified from the simulations in more detail in order of increasing level of their polymerization. Afterwards, thermodynamics of the phases and transformation mechanisms between them are discussed.

\subsection{Molecular phase $P2_1/c$}

\begin{figure}
\includegraphics[width=\columnwidth]{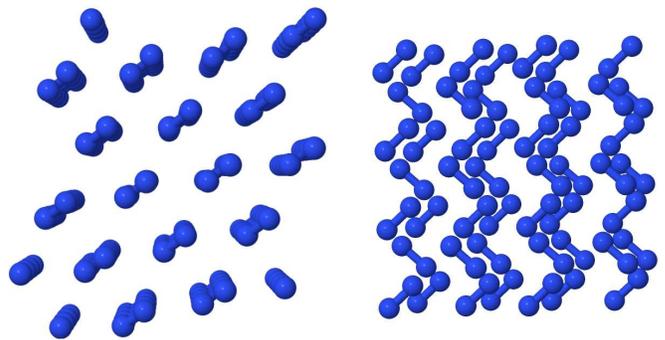}
\caption{$P2_1/c$ molecular phase at 110 GPa.}
\label{mol2}
\end{figure}

In metadynamics simulations at 110 GPa and 1300 K with $\delta s$ = 40 (kbar.\AA$^3$)$^\frac{1}{2}$ (starting from $Immm$-N$_2$), a new molecular phase emerged - Fig.~\ref{mol2}, even though this run ended up in a disordered molecular state. This phase might be interesting mainly for its low enthalpy (see discussion in section III. I), which suggests that it may represent a metastable high-pressure molecular polymorph. The spacegroup of this phase was found to be monoclinic $P2_1/c$ - complete structural data are given in Table~\ref{table} (the phase is denoted as \textit{molecular} B).

\subsection{$trans$-$cis$ chain phase}

\begin{figure}[h]
\includegraphics[width=\columnwidth]{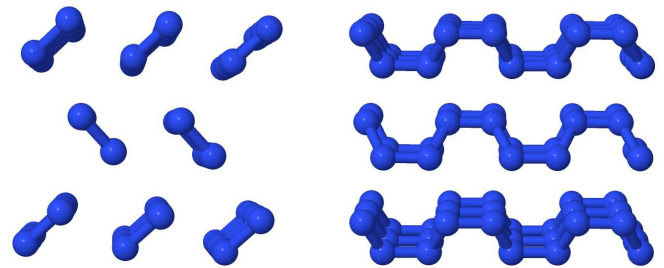}
\caption{$trans$-$cis$ chain phase at 110 GPa from two different perspectives corresponding to (001) projection (left) and to (100) projection (right) of the original simulation supercell - Fig.~\ref{MD} (which is identical to $Immm$ phase unit cell orientation - Table \ref{table}).}
\label{trans-cis}
\end{figure}

The idealized $trans$-$cis$ chain phase structure - Fig.~\ref{trans-cis}, was identified from MD and metadynamics simulations from states corresponding to the mixture of this chain form and $planar$-N phase, as described in sections III. A and B.

This phase contains only $2c$-N atoms with one shorter (double) and one longer (single) bond per atom. The N-N-N angle is around 111.6$\degree$ and N-N-N-N dihedrals are alternating in $trans = 180\degree$ and $cis = 0\degree$ conformation. This structure may be chemically viewed within a modified (or distorted) $sp^2$ hybridization scheme, where two electrons occupy one double bond, one is localized in one single bond and remaining two accommodate one electron lone-pair - for each N atom having five valence electrons (positions of lone-pairs were analyzed from the electron localization function). The structure of this phase was found to be orthorhombic $Pnma$ - for atomic positions and physical properties see Table~\ref{table}. The semimetallic character of the $trans$-$cis$ phase is consistent with observations in liquid nitrogen, where it was found that conductivity of liquid-N is strongly correlated with the amount of $2c$-N atoms representing insides of chain molecules \cite{Boates-Bonev-N-2011}.

The only previously studied phase made up of infinite $trans$-$cis$ chains is orthorhombic $ch$-phase, which was, however, found to be both mechanically (Born criteria violation) and dynamically (imaginary phonon modes) unstable up to the investigated 360 GPa \cite{Wang-N-2010}. Contrary to the $ch$-phase, our MD and metadynamics simulations indicate that the $trans$-$cis$ chain phase described in Table~\ref{table} is probably dynamically stable, which points to the importance of spatial arrangement of individual chains. Recently, a new molecular phase was predicted, which is composed of linear N$_8$ molecules with two different patterns of $trans$ and $cis$ conformations \cite{Hirshberg}. One of these molecular isomers ($trans$-$cis$-$trans$) contains a central part that is structurally identical to the pure $trans$-$cis$ chain. This N$_8$ form, which is partially polymeric (contains triple, double and single bonds as well as bonds with strong ionic character in N$_8$) and thus is likely to be also a high energy density material, was predicted to be metastable at ambient pressure and more stable than cg-N below 20 GPa.

\subsection{$2c$-$3c$ chain-like phase}

\begin{figure}
\includegraphics[width=\columnwidth]{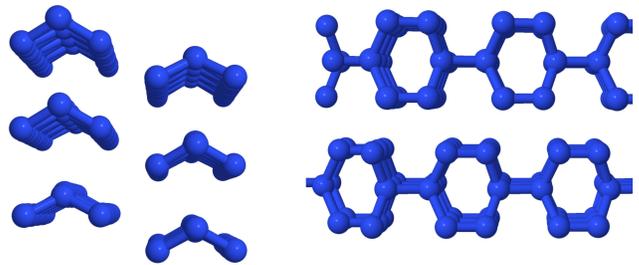}
\caption{$2c$-$3c$ chain-like phase at 110 GPa in (001) projection (left) and (010) projection (right) of the original supercell orientation.}
\label{2c-3c}
\end{figure}

In two metadynamics runs performed at 120 GPa, 1000 K with $\delta s$ = 50 (kbar.\AA$^3$)$^\frac{1}{2}$ and at 110 GPa, 1500 K with $\delta s$ = 40 (kbar.\AA$^3$)$^\frac{1}{2}$, a new type of chain-like form was created from merging of nearby $trans$-$cis$ chains into $2c$-$3c$ chains-like objects. Though these $2c$-$3c$ chains were kinetically formed in several stages of MD and metadynamics simulations (see e.g. Fig.~\ref{MD}), only in these two mentioned runs they formed a regular crystalline structure. We denote this phase as $2c$-$3c$ chain-like form - Fig.~\ref{2c-3c}, as it is composed of mixed two and three-coordinated atoms making up connected N$_6$ rings in boat conformation. Two $3c$-N and four $2c$-N atoms are present in each N$_6$ ring and the two $3c$-N atoms serve as bridges connecting the rings. The structure of $2c$-$3c$ chain-like phase was determined as monoclinic $Cm$ - see Table~\ref{table}.

\subsection{$Planar$ phase}

\begin{figure}[h]
\includegraphics[width=\columnwidth]{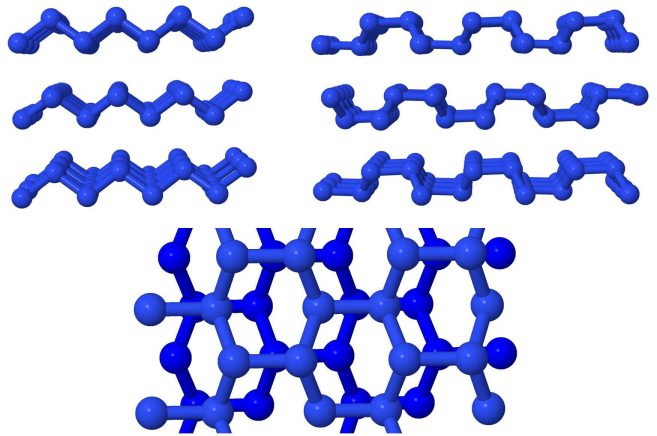}
\caption{$Planar$ phase at 110 GPa in the (001) projection (top left) and in the (100) projection (top right) of the simulation supercell illustrating how the structure may be viewed as connection of either zig-zag chains, or, alternatively $trans$-$cis$ chains. The (010) projection (bottom) shows geometry of sheets - darker blue color indicates atoms in the lower sheet.}
\label{planar}
\end{figure}

From MD at 120 GPa and 1500 K and also from 110 GPa and 1500 K metadynamics run, idealized crystalline $planar$-N structure was identified - Fig.~\ref{planar} in the same way as we identified the $trans$-$cis$ chain phase - from the structural character of the state representing their mixture. This phase is composed of fused N$_6$ rings in boat conformation, which are arranged in sheets with honeycomb geometry (when viewed along the sheet normal direction) - Fig.~\ref{planar} bottom picture. All nitrogen atoms are single-bonded and chemical bonds saturated. Sheets are mutually shifted by approximately one third of the single-bond length with respect to each other and the unit cell of $planar$-N contains eight atoms in two sheets. Structural investigation revealed spacegroup $Pnma$ with unit-cell parameters, atomic positions, density and bandgap given in Table~\ref{table}.

This structure resembles the two previously proposed layered structures - $P2_1/m$ layered boat (LB) \cite{Zahariev-N-2005} and $Pnma$ zig-zag sheet (ZS) phase \cite{Hu}, which both can be constructed by connecting either $trans$-$cis$ or zz chains. Analysis of the structure of our $planar$ phase revealed that it corresponds to the ZS form found by Hu \textit{et al.} \cite{Hu}, from which the authors built model nitrogen nanotubes.

\subsection{Transformation mechanism}

\begin{figure*}
\includegraphics[width=2\columnwidth]{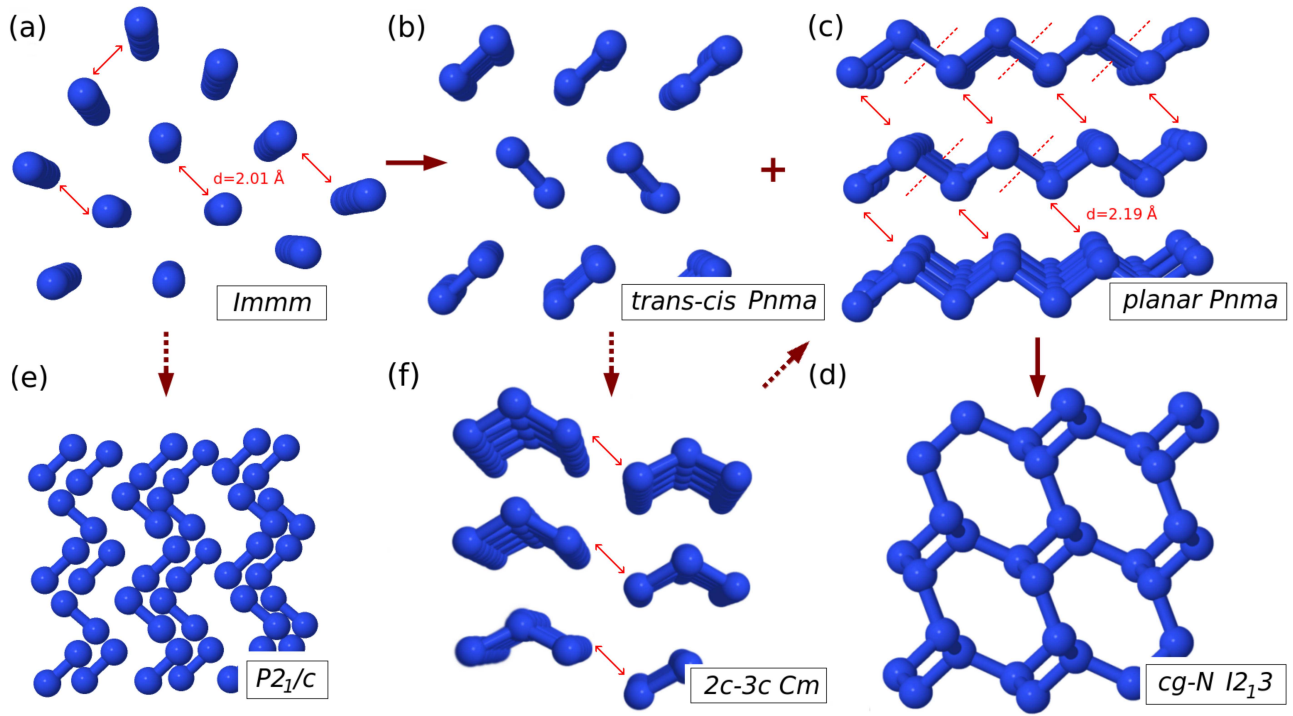}
\caption{Schematic view of the observed mechanism of molecular to cg-N structural transformation represented by (a) $\rightarrow$ (b)+(c) $\rightarrow$ (d) pathway. Red arrow lines illustrate merging of atoms, while red dashed lines indicate separations of atoms (bond breaking) and in the presented diagram are sketched only for model (c) $\rightarrow$ (d) transition. The (a) $\rightarrow$ (e) and (b) $\rightarrow$ (f) symmetry-breaking transitions were obtained outside the main pathway leading to cg-N and represent side branches of the transformation leading to two monoclinic phases: molecular $P2_1/c$ from $Immm$ and $2c$-$3c$ chain-like from the $trans$-$cis$ chain phase. The (f) $\rightarrow$ (c) transition represents a hypothetical transformation from the $2c$-$3c$ chain-like phase into the $planar$-N by simple chain merging.}
\label{mechanism}
\end{figure*}

The (a) $\rightarrow$ (b)+(c) $\rightarrow$ (d) transition pathway on Fig.~\ref{mechanism} represents the transformation mechanism from molecular $Immm$ to cg-N phase as was revealed from metadynamics simulations at 110 GPa and 1500 K. The red lines inside the pictures of the phases illustrate bond rearrangements in a simplified way. Starting from orthorhombic $Immm$ molecular phase (a) with parallel molecules, these first transform into chains with alternating $trans$ and $cis$ conformation by connecting nearby molecules, which are separated by shortest intermolecular distances of 2.01 \AA \,(calculated at 110 GPa and 0 K). During polymerization of the $Immm$ phase, two processes take place at the same time: while some molecules still merge together to form $trans$-$cis$ chains, some nearby chains connect within layers. The resulting metastable form may be described as a mixture of two phases - $trans$-$cis$ chain (b) and $planar$-N phase (c), which we denote as (b)+(c) state - see also the configurations in Fig.~\ref{meta-enth}.

\begin{figure}
\includegraphics[width=\columnwidth]{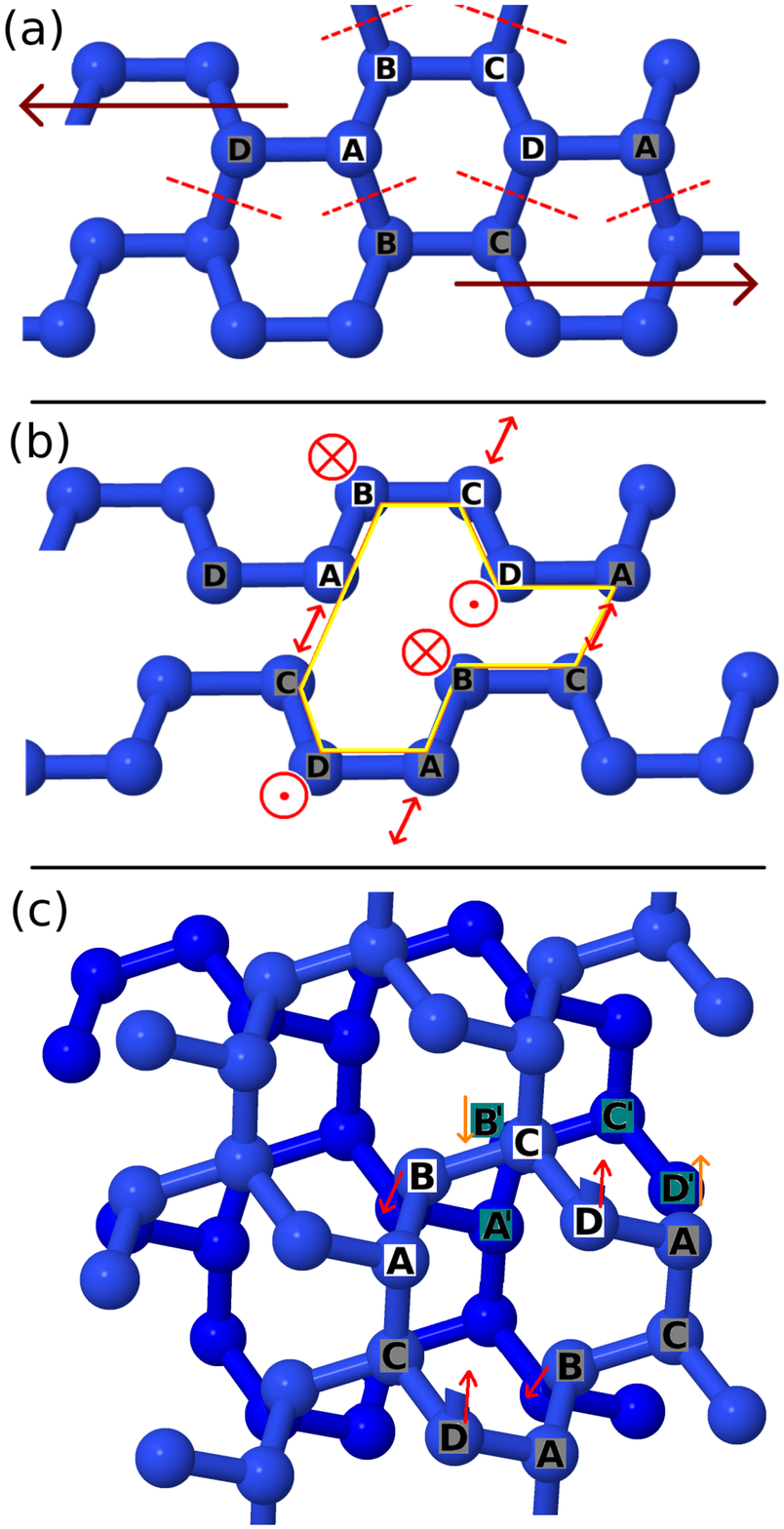}
\caption{Schematic illustration of the model transformation from $planar$-N to cg-N. (a) Inside one sheet of $planar$-N, four unit-cell atoms are chosen and labeled A-D (other atoms are shown on grey background). Each of these atoms breaks one of its bonds (A with B and C with D - as indicated by dashed lines) leaving unconnected $trans$-$cis$ chains, which then move against each other (marked by dark red arrows pointing in opposite directions). (b) After shifting, new pairs of atoms are attracted and two new bonds (per four-atoms repeating unit) are created (A with C). At this stage, connected N$_{10}$ rings emerge (one is marked by yellow contour in (b)) and two remaining atoms (B and D) are left each with one dangling bond. The very same mechanism takes place in the sheet above and in the identical sheet below. These free bonds are then saturated by connecting the separated sheets as atoms B make bonds with the sheet beneath (indicated by red vectors going into the paper) and atoms D make bonds with the sheet above (vectors going out of the paper). (c) The final structure of cg-N can be viewed as such fusion of original $planar$-N sheets, where one half of the unit-cell atoms A-D lies in one sheet and the remaining four unit-cell atoms A$^\prime$-D$^\prime$ (labeled on cyan color background) are placed in the sheet below (marked by darker blue atoms). Arrows indicate connections between the original sheets of $planar$-N.}
\label{planar-cg}
\end{figure}

The final transformation from this chain-planar mixture into cg-N (d) is illustrated on Fig.~\ref{mechanism} with red lines as a simplified process going directly from the $planar$ phase (c), though the actual one proceeds like (b)+(c) $\rightarrow$ (d). We now describe this simplified mechanism and then comment about its correspondence to the actually observed one. The (c) $\rightarrow$ (d) transformation requires chemical reformation of the whole bond network because within the sheets of pure $planar$-N all chemical bonds are already saturated and all nitrogen atoms are single-bonded. For this reason, some bonds inside the sheets must first be broken in order to properly rearrange and form cg-N. This process is schematically illustrated in more detail in Fig.~\ref{planar-cg} and may be described as follows. Dividing the process into several stages, we first focus on a single sheet formed by fused N$_6$ rings, which contains four out of eight unit-cell atoms of $planar$-N. Within this sheet, all bonds along the connected $trans$-$cis$ chains are broken and such disconnected neighboring $trans$-$cis$ chains then move in opposite directions (a). After shifting by a distance of about one bond length, new rings - N$_{10}$ are created leaving two out of four repeating atoms left with dangling bonds (b). These free bonds are thereafter allowed to make connections with surrounding sheets - one with the sheet above and one with the sheet below (c) (which first also undergo the same shifting mechanism to make unsaturated N$_{10}$ rings). This is a simplified mechanism of creation of cg-N extended network when starting from pure $planar$-N structure.

The difference between the above described model mechanism and real dynamical simulations is, that in the latter a mixture of chain and planar phases transforms into cg-N, instead of pure $planar$-N. In this state of partially saturated nitrogen, many intraplanar bonds are already broken as far as the pure $trans$-$cis$ chain phase can in fact be viewed as the $planar$ phase with all bonds between $trans$-$cis$ chains disconnected (and inversely, the $planar$ phase can be created by connection of $trans$-$cis$ chains within planes). Comparing methodologies, only the metadynamics algorithm was able to reach this final stage of the complete molecular to cg-N transformation, while in the constant-pressure MD, the system remained stuck in the mixed $trans$-$cis$ chain - $planar$-N state (energetically well-separated from cg-N) and was not able to initiate intraplanar reorganization and subsequent interplanar merging.

Alternatively to the above described mechanism, two other observed symmetry-breaking transformations from orthorhombic to monoclinic phases are shown in Fig.~\ref{mechanism} - formation of the molecular $P2_1/c$ phase from $Immm$ - (a) $\rightarrow$ (e) and of the $2c$-$3c$ chain-like phase from the $trans$-$cis$ chains - (b) $\rightarrow$ (f). The $2c$-$3c$ chains are created by connection of three nearby $trans$-$cis$ chains, which together make up two $2c$-$3c$ chains. One could speculate that the $2c$-$3c$ chain-like phase created by this process might, in principle, be ready to transform into $planar$-N straightforwardly by simply merging adjacent $2c$-$3c$ chains - (f) $\rightarrow$ (c).

The whole transformation mechanism starting from $Immm$ to cg-N may be structurally described as a progressive polymerization process governed by the $trans$-$cis$ chain geometry motif - the transformation pathway dynamically proceeds via intermediate states of more or less connected $trans$-$cis$ chains. The final cg-N form is then created by rather complicated chain reorganization, where certain bonds in sheets first break, chains shift against each other and rearrange in a way enabling interplanar bonding. We therefore propose a $trans$-$cis$ chains-based scenario as a probable candidate for the mechanism of molecular-nonmolecular transition occurring in high-pressure solid nitrogen, eventually leading to the cg-N phase.

\subsection{Enthalpies}

\begin{figure}
\includegraphics[width=\columnwidth]{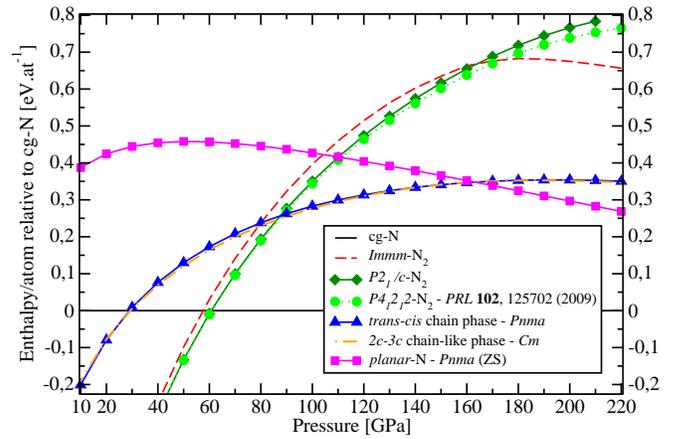}
\caption{Enthalpies versus pressure relative to cg-N for molecular phases $Immm$, $P2_1/c$ and $P42_12_12$ \cite{Pickard-Needs}, $trans$-$cis$ chain phase, $2c$-$3c$ chain-like form and $planar$-N. Note that enthalpy curves of chain phases are practically the same.}
\label{enthalpies}
\end{figure}

Calculated enthalpies of all discussed phases are presented on Fig.~\ref{enthalpies}. The starting $Immm$ phase with parallel molecules yields lower enthalpy than $P2_1/c$-N$_2$ molecular form above 160 GPa and also than phase $P42_12_12$, which was identified as the lowest-enthalpy molecular structure among those found and investigated in the study of Pickard and Needs \cite{Pickard-Needs}. This suggests that $Immm$ and also $P2_1/c$-N$_2$ may represent relatively stable high-pressure molecular polymorphs of nitrogen.

Our $trans$-$cis$ chain phase yields similar enthalpy as two zig-zag chain phases considered so far - $Cmcm$ \cite{Mattson-Sanchez-Portal} and $Imma$ \cite{Alemany-Martins} variations, at around 110 GPa (see Refs. \cite{Alemany-Martins, Kotakoski-Albe, Zahariev-N-2005, Wang-N-2010}). This suggests that the $trans$-$cis$ chain phase and the two zz-chain forms might be thermodynamically equally favored around 1 Mbar, but our simulations suggest that the $trans$-$cis$ geometry is likely to be kinetically preferred to zz chains, at least when compressing the $Immm$ molecular phase \footnote{Comparison with enthalpy of $trans$-$cis$ $ch$-phase is pointless since this form was found to be unstable \cite{Wang-N-2010}}.

From the enthalpy graph we can also see that at pressures above 160 GPa the $planar$-N phase is preferred to the chain phases and we may thus speculate that at these conditions the actual transformation mechanism proceeds as (a) $\rightarrow$ (b) $\rightarrow$ (f) $\rightarrow$ (c) $\rightarrow$ (d) (within notation of Fig.~\ref{mechanism}) instead of the described (a) $\rightarrow$ (b)+(c) $\rightarrow$ (d) pathway observed at 110 GPa. This would be also consistent with the structural comparison of the described phases, which shows that the $2c$-$3c$ form represents an intermediate structure between $trans$-$cis$ chain and $planar$ phases.

Based solely on enthalpy calculations, we find that the $trans$-$cis$ and the $2c$-$3c$ chain phase as well as the $planar$-N are highly metastable against cg-N and thus represent possible kinetic intermediate steps in the molecular to cg-N transformation. It is possible, however, that these phases might become more stable at high temperatures - a similar effect was uncovered in Ref. \cite{Erba}, where it was found that entropy of molecular forms increases upon heating faster than entropy of cg-N \cite{Erba} \footnote{Mechanism of dual role of temperature was described in this work \cite{Erba}: temperature helps kinetics to initiate polymerization, while at the same time it stabilizes molecular phases through the entropic term.}.

\section{Conclusions}

The transformation mechanism from the nitrogen molecular $Immm$ phase to the polymeric cg-N is proposed to proceed through various intermediate states sharing a common
$trans$-$cis$ chain geometry motif. Based on several metadynamics and MD simulations, we found a transformation pathway starting with formation of $trans$-$cis$ chains and further proceeding via a mixture of the $trans$-$cis$ chain phase and the $planar$ ZS phase. This configuration further reorganizes inside sheets and develops some interplanar bonds ultimately leading to the formation of the extended cg-N. Additionally, two other phases - molecular $P2_1/c$ and $2c$-$3c$ chain-like forms were obtained in separate metadynamics runs. The chain phases and the molecular $P2_1/c$ are new predictions, while the molecular $Immm$ corresponds to the B1 phase from Ref.~\cite{Hooper} and possibly also to the $Immm$ structure from Ref.~\cite{Caracas-Hemley}. The $planar$-N phase was already described as the zig-zag sheet - ZS layered structure \cite{Hu}. We found that not only can several polymeric phases be viewed as various connections of $trans$-$cis$ chains (as already recognized in Ref.~\cite{Zahariev-N-2005}), but the structure of nitrogen during the molecular-to-nonmolecular transition also dynamically evolves in a way governed by the $trans$-$cis$ chain geometry pattern. This conclusion follows from the fact that during all stages of the progressive polymerization, the individual $trans$-$cis$ chains remain stable once formed and become just distinctly connected in the intermediate states. The final step in the transformation is a collective mechanism involving the entire covalent network that transforms from layered topology of N$_6$ rings into cg-N with fused N$_{10}$ rings. Experimental observation of the transformation mechanism or some metastable phases predicted in this work would be of great interest since only cg-N and LP-N forms of polymeric nitrogen, promising high energy density material, have been synthesized so far.

\begin{acknowledgments}
This work was supported by the Slovak Research and Development Agency under Contracts No.~APVV-0558-10 and No.~APVV-0108-11 and by the project implementation
26220220004 within the Research \& Development Operational Programme funded by the ERDF. Part of the calculations were performed in the
Computing Centre of the Slovak Academy of Sciences using the supercomputing infrastructure acquired in project ITMS 26230120002 and 26210120002
(Slovak infrastructure for high-performance computing) supported by the Research \& Development Operational Programme funded by the ERDF.
\end{acknowledgments}

\bibliographystyle{aipnum4-1}

%

\end{document}